# High Precision Calibration Pairs for Southern Lucky Imaging

Matthew B. James[1], Graeme L. White[2], Roderick R. Letchford[2], Stephen G. Bosi[1].

1) University of New England, NSW, Australia.
   m.b.james27@gmail.com; sbosi@une.edu.au
2) University of Southern Queensland, QLD, Australia.
   graemewhiteau@gmail.com; Rod.Letchford@usq.edu.au

**Abstract.**

Accurate measures of double stars require accurate calibration of the instrument. Here we present a list of 50 pairs, that are quasi-evenly spaced over the southern sky, and that have Separations and Position Angles accurate at the milli-arcsec, and milli-degree level. These wide angle pairs are suggested as calibration pairs for lucky imaging observations.

## 1. Introduction.

White, Letchford and Ernest (2018) have explored the precision of historic measures with the intention of strengthening the computation of the movement of components of pairs. Such movements may include rectilinear motion or the computation of orbit parameters. That paper shows clearly that the accuracy of measurement of relative positions - separation, $\rho$, and position angle, PA - has improved over time following the changes in measurement technology.

James *et al.*, (2020b, JDSO in press) have looked at three methods of calibrating the image scale of CMOS cameras used in lucky imaging. That work compared three methods of image scale calibration for wide doubles (>5 arcsec); (i) calibration against a known pair such as α Centauri AB, (ii) the drift scan technique, and (iii) the use of a full aperture diffraction grating and narrow band filter. Details of these techniques are covered in James *et al.,* (2020b).

Of these three techniques, the α Cen AB calibration and the drift scan gave consistent results and similar formal uncertainties in the combined weighted measures of $\delta\rho \approx 80$ milli-arcsec (mas) and $\delta PA \approx 0.056°$ (James *et al*., (2020a)). The grating method will only work with narrow-band filters, such as a Hα filter with bandpass ~7nm, as filters with broader bandpass allowed the image scale to be affected by the spectrum of the star. Irrespective of the filter, the grating method resulted in an image scale that was $\approx 0.1\%$ different to those obtained with the α Cen AB and drift scan methods. On the basis of this discrepancy, and the relative simplicity of the α Cen AB (and drift scan) method we concluded that the grating and filter calibration method was unusable at the precision required.

## 2. Calibration Pairs.

The increased precision that results from the use of lucky imaging brings with it a need for corresponding improvement in image calibration.

Following the success of the α Cen AB calibration used in James *et al.*, (2019a,b and 2020a), and drawing on the results of James *et al.,* (2020b (JDSO in press)), we propose here in Table 1 a list of 50 pairs suitable for the calibration of lucky images. These pairs are quasi-evenly spaced over the Southern Hemisphere sky. They have been selected from the WDS Catalog because of their location, but also because there was little evidence of relative motion (by comparison of first and last $\rho$ and PA) between the components. In addition, all pairs are wide (between 30 and 100 arcsec) to allow accurate determination of image scale, and the stars are bright so that short exposures can be used to 'freeze out' the image translation due to atmospheric seeing – utilising the lucky imaging process.

The distribution of the pairs on the southern sky is shown in Figure 1. In the declination range $-30°$ to the equator there are 23 pairs, between $-60°$ and $-30°$ there are 15, and south of $-60°$ there are 12.

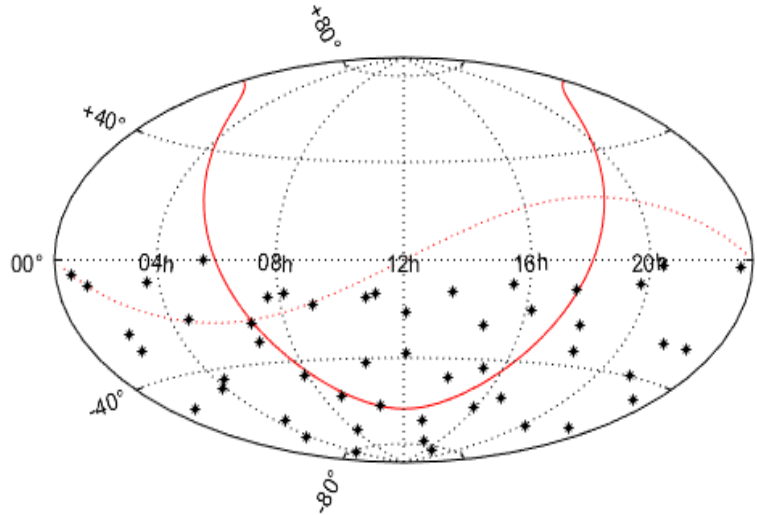

Figure 1. The distribution of calibration pairs on the southern sky.

Table 1 gives (i) the position of the primary star in J2000 coordinates, and (ii) the WDS code for the pair. The $\rho$ (iii) and PA (iv) are to allow identification 'on the night' as is the magnitudes of the primary (v) and the secondary (vi).

Table 1: Position and measures for the 50 pairs sourced from the WDS.

| # | RA (J2000) | DEC (J2000) | DISC | Epoch | PA (deg) | $\rho$ (arcsec) | Mag 1 (V) | Mag 2 (V) |
|---|---|---|---|---|---|---|---|---|
| 1 | 00 40 42.37 | -04 21 06.6 | HJ 323 | 2015 | 286 | 63.0 | 6.01 | 8.46 |
| 2 | 01 13 42.49 | -50 39 31.1 | HJ 3421 | 2010 | 81 | 62.3 | 8.49 | 8.51 |
| 3 | 01 14 24.04 | -07 55 22.2 | STFA 3 | 2018 | 331 | 47.1 | 5.19 | 7.85 |
| 4 | 02 02 44.39 | -30 19 18.6 | HJ 3478 | 2015 | 150 | 43.3 | 8.18 | 8.89 |
| 5 | 02 05 41.24 | -24 22 33.2 | I 454 | 2015 | 257 | 56.6 | 8.62 | 8.96 |
| 6 | 03 33 37.91 | -07 24 54.1 | CLL 2 | 2015 | 139 | 66.7 | 7.61 | 7.96 |
| 7 | 03 37 24.21 | -80 51 15.4 | HJ 3607 | 2015 | 128 | 35.8 | 8.47 | 9.13 |
| 8 | 04 33 33.95 | -62 49 25.2 | HJ 3670 | 2015 | 100 | 31.8 | 5.90 | 9.30 |
| 9 | 03 44 46.84 | -70 01 35.9 | LFR 1 | 2015 | 83 | 75.9 | 7.42 | 7.57 |
| 10 | 03 53 33.33 | -46 53 37.2 | SKF 949 | 2015 | 3 | 76.4 | 6.06 | 8.47 |
| 11 | 04 19 16.70 | -44 16 04.5 | HJ 3643 | 2015 | 115 | 70.3 | 5.45 | 8.64 |
| 12 | 04 39 37.02 | -21 14 51.4 | BU 1236 | 2020 | 314 | 39.9 | 7.34 | 8.96 |
| 13 | 05 38 06.52 | -00 11 03.5 | STF 758 | 2015 | 80 | 41.7 | 7.96 | 8.52 |
| 14 | 06 49 44.00 | -24 04 33.7 | S 537 | 2015 | 282 | 30.3 | 6.56 | 8.28 |
| 15 | 06 50 23.34 | -31 42 21.8 | H 5 108 | 2015 | 66 | 42.7 | 5.76 | 7.71 |
| 16 | 07 36 36.06 | -14 29 03.7 | STF1121 | 2015 | 3 | 84.0 | 6.92 | 7.66 |
| 17 | 07 49 14.30 | -46 22 23.6 | JC 11 | 1999 | 104 | 59.1 | 4.06 | 8.79 |
| 18 | 08 11 16.32 | -12 55 37.3 | SHJ 91 | 2011 | 256 | 70.1 | 4.82 | 9.37 |
| 19 | 08 19 49.00 | -71 30 54.0 | BSO 17 | 2020 | 48 | 98.4 | 5.31 | 7.67 |
| 20 | 08 56 11.24 | -55 31 42.0 | DUN 73 | 2015 | 1 | 65.9 | 7.87 | 8.29 |
| 21 | 09 03 10.05 | -17 39 38.6 | S 588 | 2012 | 327 | 32.0 | 8.61 | 8.96 |
| 22 | 10 31 13.32 | -42 13 45.6 | DUN 86 | 2017 | 291 | 83.5 | 7.34 | 8.04 |
| 23 | 10 43 09.44 | -61 10 06.7 | DUN 97 | 2015 | 17 | 83.7 | 6.59 | 8.14 |
| 24 | 10 47 37.93 | -15 15 42.9 | STF1474 | 2017 | 27 | 72.6 | 6.66 | 7.59 |
| 25 | 11 06 30.23 | -13 25 03.0 | STF1509 | 2015 | 16 | 32.9 | 7.43 | 9.36 |
| 26 | 12 04 17.11 | -21 02 21.0 | BU 458 | 2014 | 236 | 30.1 | 7.87 | 9.97 |
| 27 | 12 06 37.86 | -37 51 36.4 | HJ 4500 | 2015 | 31 | 49.9 | 6.74 | 9.07 |
| 28 | 13 15 14.94 | -67 53 40.5 | DUN 131 | 2015 | 331 | 58.2 | 4.76 | 7.24 |
| 29 | 13 32 24.72 | -12 39 47.1 | SHJ 165 | 2015 | 78 | 48.1 | 7.60 | 8.58 |
| 30 | 13 51 32.35 | -48 17 35.7 | CPO 61 | 2016 | 131 | 30.6 | 7.37 | 7.43 |
| 31 | 14 16 55.15 | -77 39 51.2 | GLI 200 | 2010 | 155 | 67.4 | 6.63 | 8.64 |
| 32 | 14 40 25.12 | -26 15 20.5 | BU 806 | 2015 | 67 | 71.4 | 7.13 | 8.92 |
| 33 | 15 10 41.61 | -43 43 47.5 | CPO 415 | 2016 | 20 | 49.7 | 7.07 | 7.66 |
| 34 | 15 28 09.57 | -09 20 49.9 | SHJ 202 | 2018 | 132 | 52.0 | 6.95 | 7.61 |
| 35 | 15 54 52.64 | -60 44 37.1 | DUN 194 | 2016 | 257 | 48.4 | 6.35 | 9.02 |
| 36 | 16 11 59.74 | -19 27 38.3 | H 5 6 | 2019 | 336 | 41.3 | 4.35 | 6.60 |
| 37 | 16 48 42.81 | -55 26 01.6 | DUN 210 | 2015 | 351 | 76.0 | 8.33 | 8.66 |
| 38 | 17 09 51.65 | -82 19 07.1 | HJ 4884 | 2010 | 5 | 34.9 | 7.15 | 8.88 |
| 39 | 17 34 46.35 | -11 14 31.2 | HJ 4964 | 2015 | 224 | 54.9 | 5.54 | 9.88 |
| 40 | 18 02 38.62 | -24 15 19.3 | ARG 31 | 2013 | 26 | 35.3 | 7.40 | 8.63 |
| 41 | 18 17 36.25 | -34 06 26.0 | HJ 5036 | 1999 | 86 | 38.9 | 6.03 | 9.48 |
| 42 | 19 43 03.26 | -08 18 26.2 | STFA 45 | 2018 | 146 | 97.0 | 7.11 | 7.56 |
| 43 | 20 14 34.92 | -64 25 46.9 | HJ 5171 | 2016 | 335 | 34.6 | 6.94 | 9.96 |
| 44 | 20 27 27.56 | -02 06 10.5 | S 749 | 2019 | 189 | 60.1 | 6.76 | 7.51 |
| 45 | 21 23 40.56 | -39 46 43.7 | SKF1173 | 2016 | 153 | 94.0 | 7.51 | 8.98 |
| 46 | 21 33 30.86 | -27 53 24.4 | BVD 137 | 2015 | 58 | 40.2 | 7.65 | 9.78 |
| 47 | 22 39 44.12 | -28 19 32.0 | H 6 119 | 2013 | 159 | 85.9 | 6.43 | 7.50 |
| 48 | 22 44 40.32 | -60 07 01.0 | HJ 5358 | 2015 | 91 | 31.2 | 8.02 | 9.95 |
| 49 | 22 58 07.06 | -46 00 44.2 | STFK1181 | 2015 | 117 | 68.7 | 8.69 | 9.42 |
| 50 | 23 28 08.47 | -02 26 53.4 | CBL 194 | 2015 | 46 | 57.4 | 8.48 | 8.11 |

### 3. Computation of Calibration ρ and PA.

Because of the necessity for the precision of the calibration ρ and PA to be better than the achievable accuracy of the observations, we have resourced the GAIA DR2 catalogue (Brown, *et al.,* (2018)), where positions and proper motions of stars are given with micro-arcsecond (µas) precision. This precision now allows the calculation of ρ and PA for any pair that Gaia has observed to a mean precision of all pairs presented in this work of ± 20 mas for $\rho$ and ± 1 milli-degree in PA at year 2100.

Equation 1 now allows the computation of the precise ρ and PA for the night of observation. This is achieved by adjusting the ρ and PA at the year 2000 by a linear increment based on the date of observation. The coefficients and constants to be used in Equation 1 are given in Table 2 and are based on the assumption that the change in ρ and PA is both small and linear between the years 2000 and 2100. Within the accuracy required the assumed linear change in ρ and PA is valid because of the choice of slow/non-movers, and their wide separation. The appendix clarifies this and presents a trigonometric and quadratic (rather than linear) equation and set of coefficients and constants.

In Equation 1 the terms are:

- $PA_t$     The position angle, PA, of the pair on the night of observation $t$.
   ($t$ = date of observation – 2000.00).
- A     The annual increment to PA. From Table 2.
- B     The PA at Year 2000.00. From Table 2.
- $ρ_t$     The separation, ρ, of the pair on the night of observation $t$.
   ($t$ = date of observation – 2000.00).
- C     The annual increment to ρ. From Table 2.
- D     The ρ at Year 2000.00. From Table 2.

***Equation 1:*** *Equations (i) PA and (ii) ρ at a particular epoch, t, for pairs listed in Table 2.*

$$PA_t = At + B \qquad \& \qquad ρ_t = Ct + D$$

### 3.1 Example of Use

For the sake of illustration, lets us assume that the observations are being made on 21$^{st}$ April 2020, using the reference pair HJ 323.

The Date (epoch) is $= 2020 + \dfrac{(31 + 29 + 31 + 21)}{365.25}$      (2020 is a leap year)

$t = 2020.307$

$t = 2020.307 - 2000$

$t = 20.307$

Obviously this calculation can be done in Excel and the day of the year (DOY) found on the web, using, for example, https://www.epochconverter.com/daynumbers.

Equation 1 now gives:

*Position Angle*

$$PA_{20.307} = -0.03222(20.307) + 286.513$$
$$= 285.859\,°$$

*Separation*

$$ρ_{20.307} = -0.01364(20.307) + 63.192$$
$$= 62.915 \; arcsec$$

For the night of April 21, 2020, the PA of HJ 323 (RA = 00 40 42.37, Dec = -04 21 06.6; Mag 1 = 6.01, Mag 2 = 8.46) is 285.859°, and the separation is 62.915 arcsec. These values are accurate to ≈ 20 mas and ≈ 0.001°.

*Table 2: Table of coefficients for the calculation of PA and ρ for the 50 pairs at a user-defined epoch for use with Equation 1.*

| | RA | DEC | Name | PA | | ρ | |
|---|---|---|---|---|---|---|---|
| | | | | A | B | C | D |
| 1 | 00 40 42.37 | -04 21 06.6 | HJ 323 | -3.222E-02 | 286.513 | -1.364E-02 | 63.192 |
| 2 | 01 13 42.49 | -50 39 31.1 | HJ 3421 | 8.986E-02 | 79.627 | 1.438E-01 | 60.820 |
| 3 | 01 14 24.04 | -07 55 22.2 | STFA 3 | -2.247E-03 | 330.863 | -5.531E-03 | 49.137 |
| 4 | 02 02 44.39 | -30 19 18.6 | HJ 3478 | 6.314E-02 | 149.537 | 2.461E-02 | 42.975 |
| 5 | 02 05 41.24 | -24 22 33.2 | I 454 | 5.638E-03 | 257.336 | 4.775E-03 | 56.548 |
| 6 | 03 33 37.91 | -07 24 54.1 | CLL 2 | 4.873E-03 | 139.029 | 9.168E-04 | 66.722 |
| 7 | 03 37 24.21 | -80 51 15.4 | HJ 3607 | 3.365E-02 | 128.261 | -1.321E-02 | 36.000 |
| 8 | 04 33 33.95 | -62 49 25.2 | HJ 3670 | 1.100E-02 | 100.016 | -1.071E-03 | 31.774 |
| 9 | 03 44 46.84 | -70 01 35.9 | LFR 1 | 1.379E-02 | 83.053 | 9.528E-04 | 75.895 |
| 10 | 03 53 33.33 | -46 53 37.2 | SKF 949 | 1.176E-02 | 3.120 | -3.708E-03 | 76.496 |
| 11 | 04 19 16.70 | -44 16 04.5 | HJ 3643 | 6.730E-03 | 115.212 | -1.487E-03 | 70.365 |
| 12 | 04 39 37.02 | -21 14 51.4 | BU 1236 | 6.174E-03 | 314.288 | 8.177E-04 | 40.417 |
| 13 | 05 38 06.52 | -00 11 03.5 | STF 758 | -1.103E-03 | 79.315 | -4.514E-04 | 41.431 |
| 14 | 06 49 44.00 | -24 04 33.7 | S 537 | 6.512E-03 | 282.319 | 1.064E-03 | 30.251 |
| 15 | 06 50 23.34 | -31 42 21.8 | H 5 108 | 7.847E-03 | 65.742 | 7.961E-06 | 42.681 |
| 16 | 07 36 35.71 | -14 29 00.3 | STF1121 | 5.146E-03 | 1.748 | -1.467E-04 | 84.293 |
| 17 | 07 49 14.30 | -46 22 23.6 | JC 11 | 4.553E-03 | 104.482 | 6.839E-03 | 59.180 |
| 18 | 08 11 16.32 | -12 55 37.3 | SHJ 91 | 9.059E-04 | 255.991 | -1.074E-02 | 69.898 |
| 19 | 08 19 49.00 | -71 30 54.0 | BSO 17 | 1.443E-02 | 47.979 | -1.681E-03 | 99.704 |
| 20 | 08 56 11.24 | -55 31 42.0 | DUN 73 | 2.236E-02 | 0.316 | -4.960E-04 | 65.883 |
| 21 | 09 03 10.05 | -17 39 38.6 | S 588 | -8.149E-03 | 326.831 | 1.228E-02 | 31.971 |
| 22 | 10 31 13.32 | -42 13 45.6 | DUN 86 | 5.032E-03 | 291.462 | -1.426E-03 | 83.411 |
| 23 | 10 43 09.44 | -61 10 06.7 | DUN 97 | 3.741E-03 | 17.372 | -3.207E-04 | 83.696 |
| 24 | 10 47 37.93 | -15 15 42.9 | STF1474 | 2.762E-02 | 26.019 | -3.068E-02 | 73.258 |
| 25 | 11 06 30.23 | -13 25 03.0 | STF1509 | 1.755E-03 | 16.106 | 1.626E-03 | 32.948 |
| 26 | 12 04 17.11 | -21 02 21.0 | BU 458 | 2.463E-03 | 233.006 | -2.253E-03 | 30.488 |
| 27 | 12 06 37.86 | -37 51 36.4 | HJ 4500 | -4.283E-04 | 30.658 | -1.157E-04 | 49.910 |
| 28 | 13 15 14.94 | -67 53 40.5 | DUN 131 | -6.384E-03 | 331.239 | -1.118E-03 | 58.220 |
| 29 | 13 32 24.72 | -12 39 47.1 | SHJ 165 | -5.228E-03 | 78.130 | 1.095E-03 | 48.097 |
| 30 | 13 51 32.35 | -48 17 35.7 | CPO 61 | -4.959E-03 | 130.624 | -3.291E-05 | 30.490 |
| 31 | 14 16 55.15 | -77 39 51.2 | GLI 200 | 3.515E-03 | 154.968 | 4.290E-03 | 67.230 |
| 32 | 14 40 25.12 | -26 15 20.5 | BU 806 | -6.072E-03 | 66.795 | -1.239E-03 | 71.413 |
| 33 | 15 10 41.61 | -43 43 47.5 | CPO 415 | -2.332E-03 | 20.431 | -5.239E-03 | 49.776 |
| 34 | 15 28 09.57 | -09 20 49.9 | SHJ 202 | -2.634E-02 | 132.807 | -5.659E-03 | 52.307 |
| 35 | 15 54 52.64 | -60 44 37.1 | DUN 194 | -1.017E-02 | 256.242 | 1.536E-05 | 48.185 |
| 36 | 16 11 59.74 | -19 27 38.3 | H 5 6 | -9.451E-03 | 335.991 | 5.603E-03 | 41.434 |
| 37 | 16 48 42.81 | -55 26 01.6 | DUN 210 | -7.384E-03 | 351.384 | 7.674E-03 | 75.913 |
| 38 | 17 09 51.65 | -82 19 07.1 | HJ 4884 | -3.574E-02 | 5.008 | -3.436E-03 | 34.909 |
| 39 | 17 34 46.35 | -11 14 31.2 | HJ 4964 | -9.218E-03 | 224.665 | 1.021E-02 | 54.678 |
| 40 | 18 02 38.62 | -24 15 19.3 | ARG 31 | 1.018E-03 | 26.307 | -1.521E-03 | 35.561 |

| 41 | 18 17 36.25 | -34 06 26.0 | HJ 5036 | -8.215E-03 | 85.128 | -1.783E-03 | 38.983 |
|---|---|---|---|---|---|---|---|
| 42 | 19 43 03.26 | -08 18 26.2 | STFA 45 | -9.619E-03 | 145.846 | -1.145E-02 | 96.766 |
| 43 | 20 14 34.92 | -64 25 46.9 | HJ 5171 | -4.404E-03 | 335.644 | 4.293E-02 | 33.650 |
| 44 | 20 27 27.56 | -02 06 10.5 | S 749 | -7.649E-03 | 188.489 | 1.044E-03 | 59.861 |
| 45 | 21 23 40.56 | -39 46 43.7 | SKF1173 | -3.479E-03 | 153.271 | 3.745E-04 | 94.382 |
| 46 | 21 33 30.86 | -27 53 24.4 | BVD 137 | -3.374E-04 | 58.024 | -2.945E-03 | 40.239 |
| 47 | 22 39 44.12 | -28 19 32.0 | H 6 119 | -2.576E-03 | 159.424 | -2.995E-03 | 86.468 |
| 48 | 22 44 40.32 | -60 07 01.0 | HJ 5358 | -4.768E-03 | 90.898 | 8.176E-05 | 31.183 |
| 49 | 22 58 07.06 | -46 00 44.2 | STFK1181 | -4.953E-03 | 117.519 | -1.066E-03 | 68.716 |
| 50 | 23 28 08.47 | -02 26 53.4 | CBL 194 | -1.804E-03 | 226.414 | -1.464E-03 | 57.393 |

**4. Limitations on Accuracy.**

The purpose of this paper is to establish a set of calibration pairs for high resolution 'lucky images' of double stars. It is noted that there are many other aspects of double star astrometry that introduce random and systematic uncertainties into measures that grow in relative importance as high precision of double star astrometry is pursued.

**4.1 Atmospheric Considerations.**

Other than the blurring of the image – commonly called seeing – due to image translation and/or defocusing, two other atmospheric effects have the potential to introduce observational bias into measures. Both develop with zenith distance.

For stars not at the zenith, atmospheric dispersion spreads star point-like images into small vertical spectra. Star light from stars of different colours are dispersed differently, so that pairs with different coloured stars will appear to be either closer or wider in their zenith distance. In practice this effect is countered by the use of coloured filters so that the light reaching the chip is of similar colour and hence dispersion. The cost is less light to the camera and longer exposures that reduce the freezing of the effects of atmosphere turbulence.

Of more importance is the atmosphere refraction which elevates the image of stars further from the zenith. The refraction depends on both the pressure and temperature of the atmosphere along the line of sight to the stars and increases as tan $z$ (where $z$ is the zenith distance). In the pair, the star farther from the zenith is elevated upwards so the vertical component of separation is measured smaller. The PA is also affected because of the distortion of the vertical measure.

Both atmospheric dispersion and refraction are well understood (Argyle, 2014 and Cox, 2002). However, it is obvious that if a large correction is required, a large degradation of the final result may occur. For this reason, precision work requires restricting the observation to a small region close to the zenith.

**4.2 Optical Considerations.**

Many of the fine optical instruments available today are of a complex optical design. Unlike the traditional long focal length refracting telescope traditionally used in double star astrometry, the focal plane of modern telescopes may have substantive aberrations that have been built into the design to allow relaxation of other parameters (such as tube length).

Of importance principally for astrometry, these aberration will affect the linearity of the image scale of the detector. That is to say, that a different separation (and PA) will be obtained using different parts of the field/chip. The most obvious of the aberrations is the field distortion (barrel or pincushion) where the focal image is either stretched, or squashed, progressively more and more moving away from the field centre.

Because the angular separation of most double stars is small, this effect is small. But with increasing accuracy being demanded, field distortion must be added to the requirement for better images. Using observations of many calibrations pairs, image distortion can be determined by centring the primary star on the chip, and plotting the pixel separation between the components against the catalogued separation. This will develop a model that can be applied to subsequent observations of targeted pairs. Importantly, this analysis should be done only for pairs close the zenith (see above).

### 4.3    Camera Considerations.

Lucky imaging works because of the high quantum efficiency and fast electronics of the modern CMOS camera. Atmospheric seeing results because of two effects; (i) the rapid translational movement of the star image in the sky (and across the chip) caused by rapidly changing prism-shaped variants in the atmosphere, and (ii) defocusing of the image caused by rapidly changing lens-like features in the atmosphere. These features change on time scales of slower than 1/1000 second, and if image exposure can be as fast as 1 kHz, then the effect of the atmosphere is 'frozen out'. For longer exposures, some effect remains but short exposures, combined with software that removes defocused images and then shift-adds the images goes close to elimination atmosphere seeing.

The technique of lucky imaging only works for images that are not saturated. Saturation is fatal for precision shift-add astrometry. To make the camera run quickly, designers have chosen to output the images in 8-bit mode, that is, the number of levels in each pixel is 256. For this reason, the dynamic range of the images is limited to about 10:1, or about 2.5 magnitudes. Pairs with large differences will struggle.

But the shortness of the exposure also depends on the gain of the camera, and subsequent photon and quantum noise. Readout noise degrades the image strongly at short exposures. The use of filters also reduces the light throughput and will lengthen the exposure, defeating the purpose of short exposures. They also reduce the effects of atmospheric diffraction. The caution here is to balance the exposure and filter requirements to obtain well exposed images that do not exceed the 256 levels of the camera.

### 5.    Summary and Conclusions.

Accurate measures of double stars require accurate calibration of the instrument. Here we present a list of milli-arcsecond accuracy calibration pairs that are quasi-evenly spaced over the southern hemisphere for use with lucky imaging observations. In the appendix we add discussion as the potential difficulties to be considered with the use of these calibration pairs.

**Acknowledgements.**

- SIMBAD Astronomical Database, operated at CDS, Strasbourg, France, https://simbad.u-strasbg.fr/simbad/
- The Aladin sky atlas developed at CDS, Strasbourg Observatory, France, https://aladin.u-strasbg.fr/
- Brian D. Mason PhD & the Washington Double Star Catalog maintained by the UNSO. (WDS), https://ad.usno.navy.mil/wds
- The GAIA Catalogue (GAIA DR2, GAIA collaboration, 2018), from VizieR, http://vizier.u-strasbg.fr/viz-bin/VizieR-3?-source=I/345/gaia2

**References.**

**Appendix: Non-linear Extrapolation.**

Equation 1, and the coefficients and constants of Table 2 above, are based on a linear extrapolation of the change of ρ and PA over a 100 years period; the years of 2000 to 2100. Two caveats are required here, (i) that the separation of the pair is dominated by the proper motion of the individual stars (with minimal orbital motion component), and (ii) that the separation is wide to minimise the non-linear projected motion.

As stated above, these potential sources of non-linear motions are avoided by choosing pairs from the WDS Catalog that have shown little relative proper motion between them. Of the 50 chosen, the mean change in ρ and PA is ≈ 1.2 arcsec and ≈ 1.4 degree respectively, over a mean time span of ≈ 155 years. Much of these changes are assumed to be in the measurement uncertainties of the observers (especially the early ones).

Wide pairs have been chosen preferentially. The separations in Table 1 range from 30.4 arcsec to 99.5 arcsec, with a mean separation of ≈ 56.5 arcsec.

To illustrate the issue or non-linear movement of close pairs, we reproduce here in Figure A1 the historic data for the optical double WDS 23238-0828 (STF 3008). Here the PA and ρ are plotted against epoch. Contrary to simplistic expectation, the true projected sky movement is rectilinear linear, and not curved in orbital motion. The strong curvature seen in Figure A1 is due to its small separation and that the historical data captured measures before and after the minimum separation.

Further details and a rectilinear analysis of this pair is given in James *et al*., (2020a).

Note: STF 3008 was chosen here for illustration – it is not one of the above 50 suggested calibration pairs.

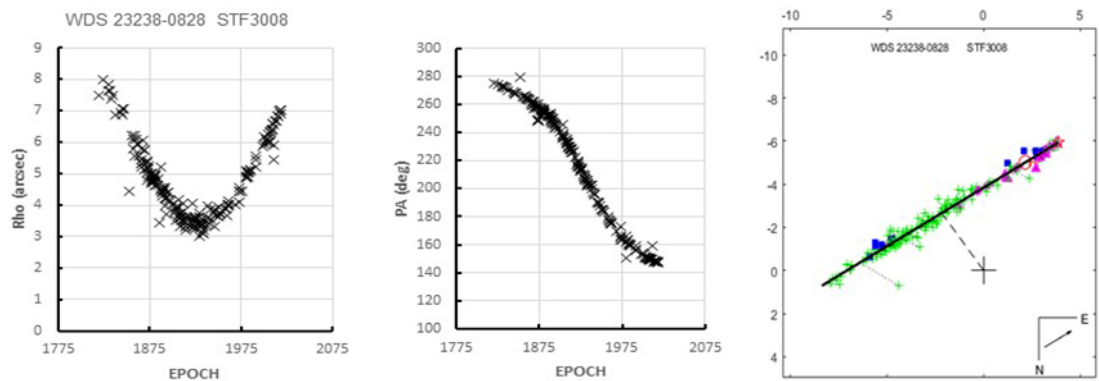

*Figure 1A: Historic data for the close optical double WDS 23238-0828 (STF 3008) showing non-linear movement of ρ and PA with epoch. The true sky motion is linear as shown in the rectilinear plot. STF 3008 is showed here for illustration only – it is not one of the above 50 suggested calibration pairs. See James et al., (2020a) for details.*

**Trigonometric and quadratic extrapolation of ρ and PA.**

Here we present inverse tangent and quadratic extrapolation for the above 50 calibration pairs. This is based on Equation A1, A2 and Table A1 and mimics the linear procedure above. Pairs denoted with an asterisk (*) require the use of Equation A1 (ii) to account for the trigonometric quadrants, and t is epoch of observation minus 2000.

*Equation A1: Equations (i) PA, (ii) PA* and (iii) ρ at an observation epoch for pairs listed in Table 2*

$$PA_t = \tan^{-1}\left(\frac{At+B}{Ct+D}\right) \quad \& \quad PA_t^* = \tan^{-1}\left(\frac{At+B}{Ct+D}\right) \pm 180 \quad \& \quad \rho_t = \sqrt{At^2 + Bt + C}$$

Equation A2: Precession *equation to account for precession in position angle.*

$$PA_p = PA_t - 0.0056\, \sin(RA_{2000})\, \sec(DEC_{2000})t$$

*Table A1: PA and ρ coefficients used for Equation A1. Items with an asterisk (*) require the use of Equation A1 (ii).*

| RA | DEC | Name | | PA | | | | ρ | | |
|---|---|---|---|---|---|---|---|---|---|---|
| | | | | A | B | C | D | A | B | C |
| 00 40 42.37 | -04 21 06.6 | HJ 323 | | 1.082E-06 | -1.683E-02 | -1.070E-05 | 4.989E-03 | 1.498E-03 | -1.855E+00 | 3.993E+03 |
| 01 13 42.49 | -50 39 31.1 | HJ 3421 | | 4.261E-05 | 1.662E-02 | -2.435E-05 | 3.042E-03 | 3.122E-02 | 1.644E+01 | 3.699E+03 |
| 01 14 24.04 | -07 55 22.2 | STFA 3 | | -8.167E-08 | -6.646E-03 | -1.808E-06 | 1.192E-02 | 4.247E-05 | -5.447E-01 | 2.414E+03 |
| 02 02 44.39 | -30 19 18.6 | HJ 3478 | * | -8.226E-06 | 6.052E-03 | -1.197E-05 | -1.029E-02 | 2.734E-03 | 1.902E+00 | 1.847E+03 |
| 02 05 41.24 | -24 22 33.2 | I 454 | * | -1.441E-06 | -1.533E-02 | 3.703E-07 | -3.444E-03 | 2.869E-05 | 5.394E-01 | 3.198E+03 |
| 03 33 37.91 | -07 24 54.1 | CLL 2 | * | 7.806E-08 | 1.215E-02 | -2.694E-07 | -1.399E-02 | 1.020E-06 | 1.223E-01 | 4.452E+03 |
| 03 37 24.21 | -80 51 15.4 | HJ 3607 | * | -3.423E-06 | 7.852E-03 | 1.592E-06 | -6.192E-03 | 1.847E-04 | -9.522E-01 | 1.296E+03 |
| 04 33 33.95 | -62 49 25.2 | HJ 3670 | * | -2.839E-07 | 8.692E-03 | 1.025E-07 | -1.535E-03 | 1.181E-06 | -6.803E-02 | 1.010E+03 |
| 03 44 46.84 | -70 01 35.9 | LFR 1 | | 2.736E-07 | 2.093E-02 | -5.750E-08 | 2.550E-03 | 1.013E-06 | 1.446E-01 | 5.760E+03 |
| 03 53 33.33 | -46 53 37.2 | SKF 949 | | 1.719E-06 | 1.156E-03 | -1.133E-06 | 2.122E-02 | 5.491E-05 | -5.715E-01 | 5.852E+03 |
| 04 19 16.70 | -44 16 04.5 | HJ 3643 | * | -3.289E-07 | 1.768E-02 | 2.714E-07 | -8.326E-03 | 2.356E-06 | -2.093E-01 | 4.951E+03 |
| 04 39 37.02 | -21 14 51.4 | BU 1236 | | -8.556E-08 | -8.037E-03 | 2.375E-07 | 7.839E-03 | 8.259E-07 | 6.608E-02 | 1.634E+03 |
| 05 38 06.52 | -00 11 03.5 | STF 758 | | -1.644E-07 | 1.131E-02 | 1.942E-07 | 2.134E-03 | 8.391E-07 | -3.747E-02 | 1.717E+03 |
| 06 49 44.00 | -24 04 33.7 | S 537 | | -2.714E-07 | -8.210E-03 | 1.428E-07 | 1.793E-03 | 1.219E-06 | 6.439E-02 | 9.151E+02 |
| 06 50 23.34 | -31 42 21.8 | H 5 108 | | 1.256E-07 | 1.081E-02 | -2.742E-07 | 4.871E-03 | 1.178E-06 | 5.618E-04 | 1.822E+03 |
| 07 36 35.71 | -14 29 00.3 | STF1121 | | -4.361E-08 | 7.143E-04 | -3.944E-08 | 2.340E-02 | 4.481E-08 | -2.474E-02 | 7.105E+03 |
| 07 49 14.30 | -46 22 23.6 | JC 11 | * | 2.028E-06 | 1.592E-02 | 2.617E-07 | -4.111E-03 | 5.418E-05 | 8.087E-01 | 3.502E+03 |
| 08 11 16.32 | -12 55 37.3 | SHJ 91 | * | 3.215E-06 | -1.884E-02 | -5.381E-07 | -4.700E-03 | 1.377E-04 | -1.504E+00 | 4.886E+03 |
| 08 19 49.00 | -71 30 54.0 | BSO 17 | | -3.358E-07 | 2.058E-02 | -3.250E-07 | 1.854E-02 | 2.831E-06 | -3.353E-01 | 9.941E+03 |
| 08 56 11.24 | -55 31 42.0 | DUN 73 | | 4.877E-06 | 1.008E-04 | -2.297E-07 | 1.830E-02 | 3.090E-04 | -9.623E-02 | 4.341E+03 |
| 09 03 10.05 | -17 39 38.6 | S 588 | | -3.502E-06 | -4.859E-03 | 1.762E-06 | 7.434E-03 | 1.992E-04 | 7.806E-01 | 1.022E+03 |
| 10 31 13.32 | -42 13 45.6 | DUN 86 | | 6.939E-07 | -2.156E-02 | 6.775E-07 | 8.477E-03 | 1.219E-05 | -2.390E-01 | 6.957E+03 |
| 10 43 09.44 | -61 10 06.7 | DUN 97 | | -4.972E-08 | 6.942E-03 | -7.778E-08 | 2.219E-02 | 1.104E-07 | -5.368E-02 | 7.005E+03 |
| 10 47 37.93 | -15 15 42.9 | STF1474 | | 4.070E-06 | 8.927E-03 | -1.169E-05 | 1.829E-02 | 1.986E-03 | -4.599E+00 | 5.367E+03 |
| 11 06 30.23 | -13 25 03.0 | STF1509 | | 1.917E-07 | 2.539E-03 | 4.147E-07 | 8.793E-03 | 2.705E-06 | 1.071E-01 | 1.086E+03 |
| 12 04 17.11 | -21 02 21.0 | BU 458 | * | 2.733E-07 | -6.764E-03 | 6.789E-07 | -5.096E-03 | 6.941E-06 | -1.376E-01 | 9.295E+02 |
| 12 06 37.86 | -37 51 36.4 | HJ 4500 | | -6.306E-08 | 7.069E-03 | 0.000E+00 | 1.193E-02 | 5.153E-08 | -1.155E-02 | 2.491E+03 |
| 13 15 14.94 | -67 53 40.5 | DUN 131 | | -2.481E-07 | -7.781E-03 | -4.911E-07 | 1.418E-02 | 3.923E-06 | -1.304E-01 | 3.390E+03 |
| 13 32 24.72 | -12 39 47.1 | SHJ 165 | | 1.522E-07 | 1.307E-02 | 7.456E-07 | 2.748E-03 | 7.504E-06 | 1.047E-01 | 2.313E+03 |
| 13 51 32.35 | -48 17 35.7 | CPO 61 | * | 9.333E-08 | 6.428E-03 | 1.231E-07 | -5.514E-03 | 3.091E-07 | -2.038E-03 | 9.296E+02 |
| 14 16 55.15 | -77 39 51.2 | GLI 200 | * | -4.938E-06 | 7.902E-03 | -3.517E-06 | -1.692E-02 | 4.764E-04 | 5.311E-01 | 4.520E+03 |
| 14 40 25.12 | -26 15 20.5 | BU 806 | | -5.994E-07 | 1.823E-02 | 5.217E-07 | 7.816E-03 | 8.184E-06 | -1.776E-01 | 5.100E+03 |
| 15 10 41.61 | -43 43 47.5 | CPO 415 | | 2.444E-07 | 4.827E-03 | -1.647E-06 | 1.296E-02 | 3.592E-05 | -5.224E-01 | 2.478E+03 |
| 15 28 09.57 | -09 20 49.9 | SHJ 202 | * | 2.501E-06 | 1.066E-02 | 5.168E-06 | -9.873E-03 | 4.272E-04 | -6.315E-01 | 2.736E+03 |
| 15 54 52.64 | -60 44 37.1 | DUN 194 | * | 1.972E-08 | -1.300E-02 | -9.833E-08 | -3.183E-03 | 1.304E-07 | 1.467E-03 | 2.322E+03 |
| 16 11 59.74 | -19 27 38.3 | H 5 6 | | -1.411E-06 | -4.683E-03 | 1.072E-06 | 1.051E-02 | 4.069E-05 | 4.634E-01 | 1.717E+03 |
| 16 48 42.81 | -55 26 01.6 | DUN 210 | | 4.022E-07 | -3.159E-03 | 2.216E-06 | 2.085E-02 | 6.571E-05 | 1.164E+00 | 5.763E+03 |
| 17 09 51.65 | -82 19 07.1 | HJ 4884 | | 7.411E-07 | 8.465E-04 | -1.027E-06 | 9.660E-03 | 2.078E-05 | -2.408E-01 | 1.219E+03 |
| 17 34 46.35 | -11 14 31.2 | HJ 4964 | * | -1.306E-06 | -1.068E-02 | -2.694E-06 | -1.080E-02 | 1.161E-04 | 1.116E+00 | 2.990E+03 |
| 18 02 38.62 | -24 15 19.3 | ARG 31 | | 9.061E-07 | 4.378E-03 | -9.278E-07 | 8.855E-03 | 2.180E-05 | -1.101E-01 | 1.265E+03 |
| 18 17 36.25 | -34 06 26.0 | HJ 5036 | | -5.181E-07 | 1.079E-02 | 2.406E-07 | 9.196E-04 | 4.228E-06 | -1.391E-01 | 1.520E+03 |
| 19 43 03.26 | -08 18 26.2 | STFA 45 | * | -4.444E-08 | 1.509E-02 | 3.822E-06 | -2.224E-02 | 1.894E-04 | -2.221E+00 | 9.364E+03 |
| 20 14 34.92 | -64 25 46.9 | HJ 5171 | | -3.853E-06 | -3.855E-03 | 1.134E-05 | 8.515E-03 | 1.858E-03 | 2.888E+00 | 1.132E+03 |
| 20 27 27.56 | -02 06 10.5 | S 749 | * | 8.747E-07 | -2.455E-03 | -4.211E-07 | -1.645E-02 | 1.221E-05 | 1.239E-01 | 3.583E+03 |
| 21 23 40.56 | -39 46 43.7 | SKF1173 | * | -3.992E-07 | 1.179E-02 | -3.169E-07 | -2.342E-02 | 3.367E-06 | 7.036E-02 | 8.908E+03 |
| 21 33 30.86 | -27 53 24.4 | BVD 137 | | -3.447E-07 | 9.481E-03 | -9.964E-07 | 5.919E-03 | 1.441E-05 | -2.376E-01 | 1.619E+03 |

| 22 39 44.12 | -28 19 32.0 | H 6 119  | * | -1.339E-07 | 8.442E-03  | 8.383E-07  | -2.249E-02 | 9.341E-06 | -5.179E-01 | 7.477E+03 |
| 22 44 40.32 | -60 07 01.0 | HJ 5358  | * | 2.528E-08  | 8.661E-03  | 1.750E-07  | -1.357E-04 | 4.052E-07 | 5.059E-03  | 9.724E+02 |
| 22 58 07.06 | -46 00 44.2 | STFK1181 | * | 1.678E-07  | 1.693E-02  | 9.681E-07  | -8.819E-03 | 1.251E-05 | -1.477E-01 | 4.722E+03 |
| 23 28 08.47 | -02 26 53.4 | CBL 194  |   | 8.592E-07  | -1.155E-02 | -3.097E-07 | -1.099E-02 | 1.081E-05 | -1.689E-01 | 3.294E+03 |